\documentstyle[12pt]{article}
\begin{document}
\begin{center}
{\large \bf
$Q^2$-evolution of 
chiral-odd twist-3 distributions\\
$h_L(x,Q^2)$
and $e(x, Q^2)$ in the large $N_c$ limit
\\ }
\vspace{5mm}
I.I. Balitsky$^1$, V.M. Braun$^{2}$, Y. Koike$^3$ 
and K. Tanaka$^4$
\\
\vspace{5mm}
{\small\it
(1) Center for Theoretical Physics, MIT,
Cambridge, MA 02139, U.S.A\\
(2) NORDITA, Blegdamsvej 17, DK-2100 Copenhagen, Denmark\\
(3) Dept. of Physics, Niigata University, Niigata 950-21, Japan\\
(4) Dept. of Physics, Juntendo University, Inba-gun, 
Chiba 270-16, Japan
\\ }
\end{center}

\begin{center}
ABSTRACT

\vspace{5mm}
\begin{minipage}{130 mm}
\small
We prove that the twist-3 chiral-odd parton distributions 
obey simple GLAP 
evolution equations in the limit $N_c\to\infty$ 
and give analytic
formulae for the corresponding anomalous dimensions. 
The results are valid to $O(1/N_{c}^2)$ accuracy and
will be useful in confronting
with future experiments.
\end{minipage}
\end{center}

The nucleon has three independent twist-3 parton
distributions $g_{2}(x, Q^{2})$, $h_{L}(x, Q^{2})$ 
and $e(x, Q^{2})$[1].
$g_{2}$ is chiral-even while $h_{L}$ and $e$ are chiral-odd.
The increasing precision of experiment 
requires understanding of these higher twist effects induced by 
correlations of partons.
In particular, $g_{2}$ and $h_{L}$ 
play a distinguished role in spin physics,
since they can be measured as leading effects for certain 
asymmetries[2,1]. 
    
The $Q^2$-evolution of twist-3 distributions is generally
quite sophisticated due to mixing with quark-antiquark-gluon operators,
the number of which increases with spin (moment of the distributions).
For the flavor-nonsinglet part of
$g_2$, a crucial simplification has been pointed out
in the large $N_c$ limit, that is neglecting $O(1/N_c^2)$
corrections[3]:
$g_2(x,Q^2)$ obeys a simple GLAP evolution equation
and the corresponding anomalous dimension is known in analytic form.

In this work[4] we demonstrate that the same pattern is
obeyed by chiral-odd distributions $h_L(x, Q^{2})$ 
and $e(x, Q^{2})$ and that the simplification 
is of universal nature.
{}For all practical purposes this solves the problem of  
the $Q^2$-evolution of twist-3 nonsinglet parton distributions,
since the corrections $1/N_c^2$ are small.

{}Following[5,1]
we define the parton distributions 
as nucleon matrix elements of nonlocal light-cone operators
(we do not show the gauge phase factors)
\begin{equation}
\int\frac{dz}{2 \pi} e^{-i (P\cdot z)x}
\langle PS|\bar\psi(z/2)\Gamma \psi(-z/2)|PS\rangle.
\end{equation}
Here $x$ is the Bjorken variable,
$z$ is a light-like vector $z^2=0$, 
and $|PS\rangle$ is the nucleon state with its momentum $P$ and 
the spin $S$.
The substitution $\Gamma = 1$ and $\sigma_{\mu \nu}i \gamma_{5}$
generates $e(x)$ and $h_{L}(x)$ respectively 
as the twist-3 contributions[1].

By using the equations of motion, 
the twist-3 
quark-antiquark distributions can be 
expressed in terms of quark-gluon
correlations. The relevant operator identities are[6] 
\begin{equation}
 \bar\psi(z)\psi(-z) = \bar\psi(0)\psi(0)
 +\int_0^1 du\int_{-u}^u dt\, S_{1}(u,t,-u)\,,
\end{equation}
\vspace{-0.5cm}
\begin{equation}
 \bar\psi(z)\sigma_{\mu\nu}z_\nu i\gamma_5\psi(-z) =        
\big[\bar\psi(z)\sigma_{\mu\nu}z_\nu i\gamma_5\psi(-z)]_{\rm twist 2}
+i z_\mu \int_0^1 \!\!\! udu\int_{-u}^u \!\!\!
tdt\,S_{i \gamma_{5}}(u,t,-u),
\end{equation}
where 
$S_{\Lambda}(u,t,v;\mu^2) = \bar \psi(u z)\sigma_{\mu\xi}\Lambda 
   g G_{\nu\xi}(tz)z_\mu z_\nu \psi(vz)$
with $\Lambda = 1, i\gamma_{5}$,
and we neglect operators containing total 
derivatives which are irrelevant for our purposes.

The $Q^2$-dependence of the twist-3 distributions is governed by the
renormalization group (RG) equation for the corresponding nonlocal operators
$S_{\Lambda}$. 
To leading logarithmic accuracy the 
evolution for $\Lambda=1$ and $i \gamma_{5}$ is the same; hence we
drop the subscript in what follows.
We introduce the Mellin transformed operators[7,3]
\begin{equation}
     S(u,t,v)=
     \frac{1}{2\pi i}\int^{1/2+i\infty}_{1/2-i\infty}\!\!dj\,
      (u-v)^{j-2}S(j,\xi), \hspace{0.5cm}
      \xi=\frac{u+v-2t}{u-v},
\end{equation}
with $j$ the complex angular momentum; operators with different $j$ 
do not mix with each other. Neglecting contributions 
down by $1/N_c^2$, we obtain 
the RG equation[4]
\begin{equation}
     \left(\mu \frac{\partial}{\partial\mu} 
           + \beta(g)\frac{\partial}{\partial g}\right)
      S(j,\xi;\mu)= -\frac{\alpha_{s}}{2\pi}
   \int_{-1}^{1}\!\!d\eta \,K_{j}(\xi,\eta)
 S(j,\eta;\mu).
\end{equation}
{}For the explicit form of the kernel $K_{j}(\xi,\eta)$,
we refer the readers to Ref.[4].

To solve (5) we consider the  {\em conjugate} 
homogeneous equation
\begin{equation}
   \int_{-1}^{1}\!\!d\eta \,K_{j}(\eta,\xi) \phi_{j}(\eta)
=     \gamma_{j}\phi_{j}(\xi).
\end{equation}
Then it is easy to see that 
$\int^{1}_{-1} d\xi \phi_{j}(\xi) S(j, \xi; \mu)$
gives a multiplicatively
renormalizable {\em nonlocal} operator corresponding to 
the anomalous dimension 
$\gamma_j$.

One can prove that Eq.(6) has two solutions 
analytic 
at the points $\xi = \pm 1$:
$\phi^{+}(\xi)=1$ and 
$\phi^{-}(\xi)=\xi$
(all other solutions have logarithmic branching points).
The corresponding eigenvalues (anomalous dimensions) 
respectively equal
\begin{equation}
     \gamma_{j}^{\pm} =
2 N_c\left\{\psi(j+1)+\gamma_{E}-\frac{1}{4}-\frac{1}{j+1}
\left(- \frac{1}{2} \pm 1 \right) \right\},
\end{equation}
where $\psi(z)=\frac{d}{dz}\ln\Gamma(z)$
and $\gamma_{E}$ is the Euler constant.

The superscript $\pm$ corresponds to the ``parity''
under $\xi \rightarrow - \xi$: due to the 
symmetry of the kernel $K_{j} (-\eta, -\xi)= K_{j} (\eta, \xi)$[4], 
one can look for separate solutions 
which are even (odd) under $\xi\to -\xi$. 
{}From Eqs.(2) and (3), 
the relevant quantities for $e(x)$ and $h_L(x)$ are
even and odd ``$\xi$-parity'' 
pieces of the nonlocal operators.

Substituting the definition (4) into (2) 
and (3), 
we observe that our solutions
in the large $N_{c}$ limit give the 
$Q^{2}$-evolution for the moments:
\begin{equation}
  {\cal M}_n[e](Q) = L^{\gamma^+_n/b}{\cal M}_n[e](\mu)\,;
\;\;
  {\cal M}_n[\widetilde{h}_L](Q) = L^{\gamma^-_n/b}
{\cal M}_n[\widetilde{h}_L](\mu),
\end{equation}
where 
${\cal M}_n[e]\equiv \int_{-1}^1 dx x^n e(x)$, 
$b = (11N_c-2N_f)/3$,
and 
$L\equiv \alpha_s(Q)/\alpha_s(\mu)$.
$\widetilde{h}_L$ is the genuine twist-3 contribution to $h_{L}$,
after subtracting out the twist-2 piece[1].
These results show simple GLAP evolutions
without any complicated operator mixing.

Expansion of nonlocal operators at small quark-antiquark separations
generates the series of local operators of increasing dimension.
The anomalous dimension matrix for these local operators
has been obtained for $h_{L}$[8] as well as for $e$[9].
By comparing our solutions (7) with the spectrum of the anomalous 
dimensions in the $N_{c} \rightarrow \infty$ limit,
which is obtained by the numerical diagonalization
of the mixing matrix in [8,9], we conclude that our 
solutions always correspond to operators with the {\em lowest}
anomalous dimension in the spectrum (for the detail, see Refs.[4,9]). 


%
%
%

To illustrate numerical accuracy of the leading-$N_c$ approximation,
consider the result[8] including $1/N_c^2$ corrections 
for the evolution of the $n=5$ moment of $h_L$:
\begin{eqnarray}
{\cal M}_{5}[\widetilde{h}_{L}](Q)
&=& \left[ 0.416 b_{5,2}(\mu) + 0.193 b_{5,3}(\mu) \right]
L^{12.91/b}
\nonumber \\
&+& \left[ 0.013 b_{5,2}(\mu) - 0.050 b_{5,3}(\mu) \right]
L^{18.05/b},
\end{eqnarray}
where $b_{n, k}(\mu)$ 
are reduced matrix 
elements of the independent 
quark-antiquark-gluon local operators in the notation of [1,8].
This is reduced in the large $N_c$ limit to
\begin{equation}
 {\cal M}_{5}[\widetilde{h}_{L}](Q) = \left[
\frac{3}{7}b_{5,2}(\mu) + \frac{1}{7} b_{5,3}(\mu) \right]
L^{13.7/b}.
\end{equation}
One observes: the contribution of the 
operator with the higher anomalous dimension in (9) is small
($\sim 1/N_{c}^{2}$), while the one with
the lowest anomalous dimension is close to the large $N_{c}$ limit.
This observation is crucial for phenomenology, 
since description of each 
moment of the twist-3 distribution now 
requires one single nonperturbative
parameter. Similar phenomenon is observed also for $e(x)$[9].


To summarize,  our solutions
provide a powerful framework both in confronting with 
experimental data and for the model-building. From a general
point of view, they are interesting as providing with 
an example of an interacting
three-particle system in which one can find an exact energy 
of the lowest state. For phenomenology, main lesson is that inclusive
measurements of twist-3 distributions are complete (to our accuracy)
in the sense that knowledge of the distribution at one value of $Q_0^2$ 
is enough to predict its value at arbitrary $Q^2$, in the spirit
of GLAP evolution equation. 

\vspace{0.2cm}
\vfill
{\small\begin{description}
\item{}
This work is financially supported in part by RIKEN.
\item{[1]} 
R.L. Jaffe and X. Ji,
Nucl. Phys. {\bf B375} (1992) 527.
\item{[2]} 
K. Abe et al., 
Phys. Rev. Lett. {\bf 76} (1996) 587.
\item{[3]} 
A. Ali, V.M. Braun and G. Hiller,
Phys. Lett. {\bf B266} (1991) 117.
\item{[4]} 
I.I. Balitsky, V.M. Braun, Y. Koike and K. Tanaka,
Phys. Rev. Lett. {\bf 77} (1996) 3078.
\item{[5]} 
J.C. Collins and D.E. Soper,
Nucl. Phys. {\bf B194} (1982) 445.
\item{[6]} 
V.M. Braun and I.E. Filyanov,
Z. Phys. {\bf C48} (1990) 239.
\item{[7]} 
I.I. Balitsky and V.M. Braun,
Nucl. Phys. {\bf B311} (1988/89) 541.
\item{[8]} 
Y. Koike and K. Tanaka, 
Phys. Rev {\bf D51} (1995) 6125.
\item{[9]} 
Y. Koike and N. Nishiyama,
in these proceedings; hep-ph/9609207.
\end{description}}

\end{document}